\documentclass[10pt]{WileyMSP-template}
\usepackage{siunitx}
\usepackage[english]{babel}
\usepackage{amsmath,amssymb,mathrsfs}
\usepackage{psfrag}
\usepackage{graphicx}
\usepackage{graphics}
\usepackage{epsfig}
\usepackage{bm}
\usepackage{color}
\usepackage{verbatim,color}
\usepackage{physics}
\usepackage[normalem]{ulem}
\usepackage{cancel}
\usepackage[colorlinks=true,linkcolor=blue,citecolor=black,urlcolor=blue]{hyperref}

\usepackage{subcaption}

\usepackage[acronym]{glossaries}
\newacronym{dr}{DR}{dimensional regularization}
\newacronym{ms}{MS}{minimal subtraction}
\newacronym{eft}{EFT}{effective field theory}
\newacronym{mqst}{MQST}{macroscopic quantum self-trapping }
\newacronym{lhy}{LHY}{Lee-Huang-Yang}
\newcommand{\beq}{\begin{equation}}
\newcommand{\eeq}{\end{equation}}
\newcommand{\beqa}{\begin{eqnarray}}
\newcommand{\eeqa}{\end{eqnarray}}
\newcommand{\ba}{\begin{aligned}[b]}
\newcommand{\ea}{\end{aligned}}

\definecolor{darkgreen}{rgb}{0.0, 0.5, 0.0}
\definecolor{shadow}{rgb}{0.6, 0.6, 0.6}
\definecolor{darkgreen}{rgb}{0.0, 0.5, 0.0}
\definecolor{c1}{rgb}{0.9, 0.0, 0.0}
\definecolor{c2}{rgb}{0.3, 0.0, 0.6}
\definecolor{c3}{rgb}{0.9, 0.0, 0.0}

\newcommand{\Ra}[1]{\color{black} #1 \color{black}  }

\begin{document}

\pagestyle{fancy}
\rhead{\vspace{0.5cm}}

\title{Low-energy atomic scattering: s-wave relation between the interaction potential and the phase shift}
\maketitle

\author{Francesco Lorenzi$^{1, 2,*}$}
\author{Luca Salasnich$^{1,2,3}$}

\dedication{}

\begin{affiliations}
$^{1}$ Dipartimento di Fisica e Astronomia ``Galileo Galilei", Università di Padova, Via Marzolo 8, 35131 Padova, Italy\\
$^{2}$  Istituto Nazionale di Fisica Nucleare (INFN), Sezione di Padova, via Marzolo 8, 35131 Padova, Italy\\
$^{3}$Istituto Nazionale di Ottica (INO) del Consiglio Nazionale delle Ricerche (CNR), via Nello Carrara 1, 50019 Sesto Fiorentino, Italy \\
*Email Address: \texttt{francesco.lorenzi.2@phd.unipd.it}
\end{affiliations}


\begin{abstract}

We investigate the on-shell approximation in the context of s-wave scattering for ultracold two-body collisions. Our analysis systematically covers spatial dimensions $D=1,2,3$, with the aim of identifying the regimes in which the approximation remains valid when applied to commonly used model interaction potentials. Specifically, we focus on the square well and delta shell potentials, both of which admit analytical solutions for the s-wave scattering problem in all dimensions considered.
By employing the exact analytical expressions for the s-wave scattering phase shift, we perform a direct comparison between the exact on-shell matrix element of the interaction potential and their corresponding approximations across a range of collision momenta. Particular attention is given to the low-energy regime. Our findings indicate that, although the on-shell approximation generally improves with increasing momentum, its accuracy also improves for weaker potentials.
Remarkably, in the limit of weak interactions, we demonstrate that the on-shell approximation becomes exact at leading order. In this regime, the approximation offers a controlled means of deriving the low-momentum expansion of the potential and may serve as a useful tool in constructing effective interactions for quantum field theories.
\end{abstract}

\section{Introduction}
Low temperature behavior of trapped gases is characterized by the interaction between atoms \cite{stoof_ultracold_2010}, which has been studied extensively, especially since in the experiments it can be tuned by using Feshbach resonances \cite{pethick_boseeinstein_2008} over a wide range of values. 
These effects translate into a nonlinear effective self-interaction in the mean-field picture, that can be tuned from attractive to repulsive, allowing the study and direct observation of nonlinear structures like matter-wave solitons \cite{khaykovich2002formation}.
The remarkable configurability of the atomic traps allows the study of low-dimensional systems \cite{bardin_quantum_2024}, in particular planar (two-dimensional) \cite{adhikari_quantum_1986}, linear (one-dimensional) \cite{barlette2000quantum}, and lately also systems with nontrivial curvature \cite{tononi_low-dimensional_2023}.
The two-body interaction physics gets modified when considering reduced dimensionality, and while the parameters available from the experiments pertain to low-energy scattering \cite{egorov_measurement_2013, bartenstein_precise_2005,inouye_observation_1998}, their inclusion in the theory presents some subtleties.

\hspace{0.2cm} %
Field theories addressing the many-body physics of ultracold gases can implement potentials of the contact type, in some formulations including derivatives of the Dirac delta to model the shape of the interaction potential beyond the universal regime, which is the regime where the s-wave scattering length is the only relevant parameter \cite{cappellaro_thermal_2017, planasdemunt2024equation}.
The couplings are usually determined expanding the interaction potential for low momentum and matching the expansion to reproduce the low-energy scattering observables \cite{roth_effective_2001, castin_bose-einstein_2001}. In a recent study, we proposed an approximate method to achieve this mapping in all dimensions \cite{lorenzi_-shell_2023}. This method, called on-shell (OS) approximation was applied to the derivation of the two-dimensional equation of state for a Bose gas in presence of effective range correction. 

\hspace{0.2cm} %
In this article, we address the problem of determining the validity of the OS approximation in the case of commonly adopted potentials in ultracold atomic scattering. Since the OS approximation provides an equality between the s-wave component of the interaction potential and a function of the s-wave scattering phase shift, as functions of the momentum $k$, we check directly the validity for all $k$, focusing especially in the low-momentum regime of the approximation. 
We tackle potentials for which exact inverse scattering solutions are nontrivial in methods like the Gel'fand Levitan Marchenko equation \cite{newton_scattering_2013, rakityansky_jost_2023, taylor_scattering_2012}, but which at the same time are treatable solving the Schr\"odinger equation in all dimensions. We thus consider potentials of the square well and the delta shell types.

\hspace{0.2cm} %
The article is organized as follows: in Section 2 we recall the main results of the OS approximation, and provide a unified formula for the relation between the Fourier transform of a potential and its s-wave on-shell component; in Section 3 we provide a verification of the OS approximation in the case of the square well potential and the delta shell potential, by solving analytically the scattering problem and evaluating the approximation errors. 
In section 4 we analyze the results and draw a general conclusion for the validity of OS approximation. 
In the Appendix A we review the expressions of Fourier transform in component of the potential to the Fourier transform, and in Appendix B we show the Born approximation in all dimensions as a reference.

\section{Phase shift and interaction potential in the on-shell approximation}
The fundamental quantity for scattering of ultralow-temperature atoms is the s-wave scattering phase shift $\delta_0(k)$ as a function of $k$, the modulus of the momentum in the relative motion reference frame \cite{pethick_boseeinstein_2008}. In particular, its behavior for $k\to 0$ is used to define the s-wave scattering length for any dimension $D$, fixing the universal behavior at low densities. 
In our past treatment of the low-energy two-body scattering for ultracold atoms, we provided an approximation (denoted as OS approximation) for the integral appearing in the Lippman-Schwinger equation for the $T$-matrix \cite{lorenzi_-shell_2023}.
This allowed to express of the s-wave and on-shell component of the interaction potential matrix $V_0(k)$ by means of a simple algebraic equation, readily solved in terms of the scattering phase shift in s-wave $\delta_0(k)$ in the following way
\begin{equation}\label{eq:osa}
     V_0(k) \overset{\mathrm{O S}}{=} \begin{cases}
     - \dfrac{2 \hbar^2}{m} k\tan(\delta_0(k) ) \,, \qquad &D=1\\
     -  \dfrac{4 \hbar^2}{m} \left( \cot(\delta_0(k))- \dfrac{2}{\pi} \ln\left( \dfrac{k}{2} \dfrac{e^{\gamma_E/2}}{\Lambda}\right) \right)^{-1} \,, \qquad  &D=2 \\
     - \dfrac{4\pi \hbar^2}{m} k^{-1} \ \tan(\delta_0(k)) \,, \qquad &D=3
     \end{cases}
\end{equation}
where $\hbar$ is the reduced Planck constant, and $m$ is twice the \Ra{reduced mass of the system of the two colliding atoms, defined as $m =2 m_Am_B/(m_A+m_B)$ with $m_A$ and $m_B$ the masses of the two atoms.}
In the derivation of the above relations, dimensional regularization was used to compute the ultraviolet divergent integral of the Lippman-Schwinger equation, appearing in all dimensions. In $D=2$, minimal subtraction was also utilized, and the momentum cutoff $\Lambda$ was introduced as a regularizer. \Ra{In the same equation, $\gamma_E$ is the Euler-Mascheroni constant.}
We remark that the OS approximation bears some structural similarities to the Born approximation for the s-wave scattering (see the Appendix B) in the fact that both are able to link the potential to the scattering phase shift. However, unlike the OS approximation, the Born approximation is giving, in principle, a complex value for the potential, a result that must be interpreted case by case.
In Ref. \cite{lorenzi_-shell_2023}, we have shown that Eq.~(\ref{eq:osa}) is crucial to determine a meaningful relationship between the low-energy scattering parameters, i.e. the s-wave scattering length $a_s$ and s-wave effective range $r_s$ \cite{macedo-lima_scattering_2023, arnecke_effective-range_2006, jeszenszki_s-wave_2018}, and the parameters $\tilde{g}_0$ and $\tilde{g}_2$ of the expansion $\tilde{V}(k) = \tilde{g}_0+\tilde{g}_2k^2+\dots$ involving the Fourier transform $\tilde{V}(k)$ of the interaction potential. 
This result can be useful to obtain finite range corrections to the usual effective potentials in the quantum field theory of ultracold atoms \cite{fu_beyond_2003, veksler_simple_2014, zinner_effective_2012, pendse_probing_2018}.
\Ra{The relationship is built using the so-called effective range expansion, as this expansion links the phase shift $\delta_0(k)$ to the experimentally accessible parameters of scattering length $a_s$ and effective range $r_s$. The expansion is dependent on the dimension \cite{bethe_theory_1949, schw, lorenzi_-shell_2023} in the following form
\begin{align}
k \ \tan({\delta_0(k)}) &= {1\over a_s} + {1\over 2} r_s k^2 + ... \,, && D=1\\
\cot(\delta_0(k)) &= {2\over \pi} \ln({k\over 2} a_s e^{\gamma_E}) 
+ {1\over 2} r_s^2 k^2 + ... \,,  && D=2 \\
k \ \cot({\delta_0(k)}) &= - {1\over a_s} + {1\over 2} r_s k^2 + ...   \,,           && D=3
\end{align}
We remark that in the $D=3$ case,  a different kind of expansion, valid for low scattering length, is discussed in Ref.~\cite{Adhikari_2018}.
}

\hspace{0.2cm} %
The definition of the phase shift $\delta_0(k)$ is easily found in the literature \cite{sakurai_modern_1985, adhikari_quantum_1986, barlette2000quantum}, but the computation of $\delta_0(k)$ can be quite demanding. Instead, the link between the s-wave component $V_0(k)$ of an interaction potential $V(r)$ and its Fourier transform $\tilde{V}(k)$ presents several subtleties.
The s-wave component $V_0(k)$ is computed with the s-wave projection of the function $V(r)$ as follows \cite{adhikari_effective_1982, barlette2001integral, sakurai_modern_1985}
\begin{equation}\label{eq:swave-def}
    V_0(k) = S_D \int_0^\infty dr \, r^{D-1} \ V(r) \ \Phi_D(kx)^2 \,,
\end{equation}
where $S_D=2 \pi^{D/2}/\Gamma(D/2)$ is the $D$-dimensional spherical surface, taking values $S_1 = 2$, $S_2 = 2\pi$ and $S_3 = 4\pi$; and the basis functions $\Phi$ are 
\begin{equation}
     \Phi_D(x) = \begin{cases}
    \cos(x) \,, \qquad &D=1 \\
     J_0(x) \,, \qquad &D=2 \\
     j_0(x) \,, \qquad &D=3
     \end{cases}
\end{equation}
where $J_0$ is the zeroth-order Bessel function, and $j_0$ is the zeroth-order Riccati Bessel function (also known as spherical Bessel function) \cite{abramowitz_handbook_1965, newton_scattering_2013, rakityansky_jost_2023}.
We would like to connect the function $V_0(k)$ to the more well-known $D$-dimensional Fourier transform of the potential, which is
\begin{equation}\label{eq:fourier}
\tilde{V}(\mathbf{q}) = \int d^D\mathbf{r}\, e^{i\mathbf{q} \cdot \mathbf{r}} \, V(r) \,.
\end{equation}
In scattering theory, the Fourier transform $\tilde{V}(\mathbf{q})$ is calculated with $\mathbf{q} = \mathbf{k}-\mathbf{k}'$ representing the momentum difference between an incoming state with momentum $\mathbf{k}$ and an outgoing state $\mathbf{k}'$.
Since the interaction potential $V(r)$ we are considering is radially symmetric, the function $\tilde{V}(\mathbf{k}-\mathbf{k}')$ only depends on the modulus of the momentum difference $|\mathbf{k}-\mathbf{k}'|$.
In the case of $|\mathbf{k}| = |\mathbf{k}'| = k$, corresponding to the on-shell condition, the momentum difference can be written as $|\mathbf{k}-\mathbf{k}'| = 2k\sin(\phi/2)$, $\phi$ being the angle between the two wavevectors. The angle $\phi$ may be considered as a generalized angle in $D$ dimensions, taking values in the following sets or ranges:  $\phi\in\{0, \pi\}$ for $D=1$, $\phi\in[0, 2\pi]$ for $D=2$, and $\phi\in[0, \pi]$ for $D=3$. 
%
Selecting the s-wave component of the potential is equivalent to considering only the isotropic scattering in the whole space. The corresponding scattering amplitude is required to be symmetric with respect to parity in $D=1$, circularly symmetric in $D=2$, and radially symmetric in $D=3$.
This symmetry requirement suggest that the s-wave component can be extracted by using averages of the Fourier transform $\tilde{V}(\mathbf{k}-\mathbf{k}')$ over the angle $\phi$. 
These averages can be carried out, respectively, over the forward and backward directions in $D=1$, over the circle in $D=2$, and the spherical surface in $D=3$,
\begin{equation}\label{eq:swave-vs-fourier}
    V_0(k) = \begin{cases} 
    \frac{1}{2}\left(\tilde{V}(0)+\tilde{V}(2k)\right) \,, \qquad &D=1\\[1.5ex]\displaystyle
    \frac{1}{2\pi} \int_0^{2\pi} d\phi\, \tilde{V}\left(2k\sin(\phi/2)\right) \,, \qquad &D=2\\[1.5ex]\displaystyle
    \frac{1}{2} \int_{0}^\pi d\phi \, \sin(\phi)  \, \tilde{V}(2k\sin(\phi/2)) \,, \qquad &D=3
    \end{cases}
\end{equation}
It is possible to derive the relation (\ref{eq:swave-def}) by noting the following.
The $D=1$ expression in Eq.~(\ref{eq:swave-vs-fourier}) simply evaluates two points of the Fourier transform at opposite angles. By substituting the definition (\ref{eq:fourier}), we notice that the two terms $\tilde{V}(0)$ and $\tilde{V}(2k)$ are expressed as the plain integral of $V(r)$ on all the space, and the integral of  $V(r)\cos(2kr)$, respectively. The s-wave expression for is retrieved using the duplication formula for the cosine $\cos^2(x) =(1+\cos(2x))/2 $. The cases of $D=2$ and $D=3$ are more similar.
In $D=2$ angular integral simplifies due to the following identity \cite[6.519]{gradshteyn_table_2014}
\begin{equation}\label{eq:J}
\frac{1}{2\pi} \int_0^{2\pi} d\phi\, J_0\left(\sqrt{2k^2(1-\cos\phi)}r\right) = J_0(kr)^2 \,,
\end{equation}
Similarly, in the $D=3$ case the angular integral is computed using the identity \cite{abramowitz_handbook_1965}
\begin{equation}\label{eq:j}
\frac{1}{2} \int_{0}^{\pi} d\phi \, \sin(\phi)\, \frac{\sin\left(\sqrt{2k^2(1-\cos\phi)}r\right)}{\sqrt{2k^2(1-\cos\phi)}r} = \left(\frac{\sin(kr)}{kr}\right)^2 \,.
\end{equation}
Combining Eqs. (\ref{eq:J}, \ref{eq:j}) with the $D=2, 3$ expressions in Eq.~(\ref{eq:swave-vs-fourier}) and the definition (\ref{eq:fourier}) of the Fourier transform, the result (\ref{eq:swave-def}) is retrieved (see also Appendix A).
A peculiar feature of the relation (\ref{eq:swave-vs-fourier}) is that the low-momentum expansion of the s-wave on-shell component and the Fourier transform are not coincident, namely, if $V_0(k)\sim g_0+g_2 k^2$, and $\tilde{V}(k) \sim \tilde{g}_0+\tilde{g}_2k^2$ for $k \to 0$, we have $g_0=\tilde{g}_0$ and $g_2=2\tilde{g}_2$.

\section{Exact results for model potentials}

\hspace{0.2cm}%
In the following, we report some examples of potentials for which analytic solutions exist in all dimensions. The two-body Schr\"odinger equation is considered in its s-wave radial part, and the stationary scattering problem can be solved, namely, the phase shift is determined as a function of the parameters defining the potential. This information is used to verify the OS approximation. We investigate the square well and the delta shell potentials, in the regime of weak attraction and repulsion, which is the one of interest for the case of potentials which do not develop bound states.

\subsection{Square well}
\begin{figure}[ht]
  \centering
  \begin{subfigure}[b]{0.42\textwidth}
    \includegraphics[width=\textwidth]{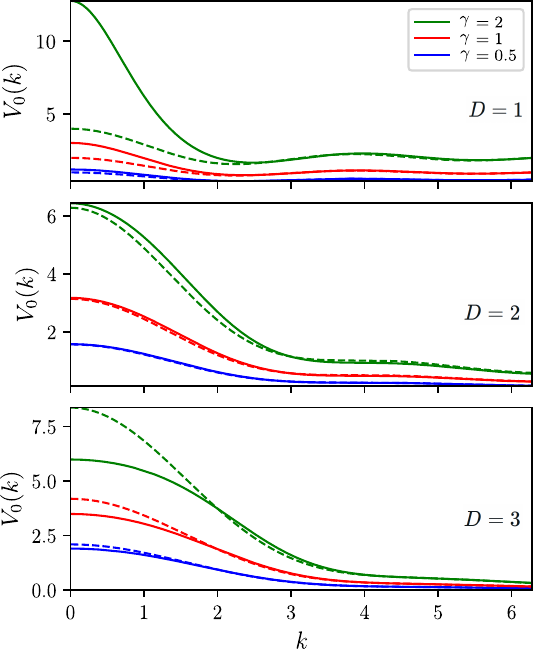}
    \caption{Repulsive square well.}
  \end{subfigure}
  \hfill
  \begin{subfigure}[b]{0.42\textwidth}
    \includegraphics[width=\textwidth]{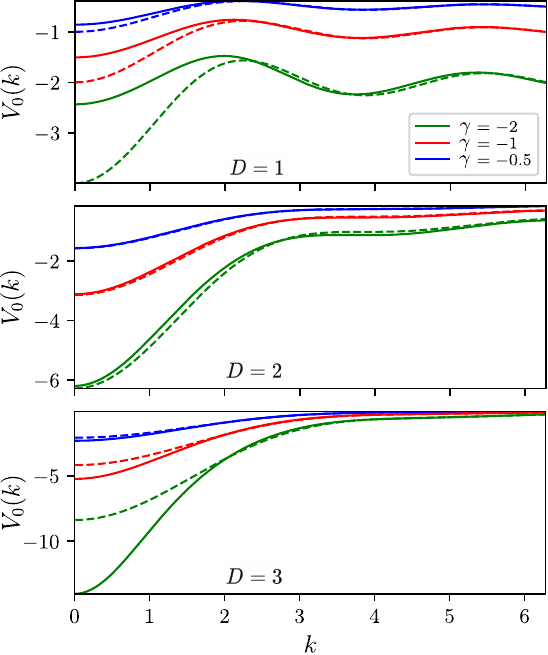}
    \caption{Attractive square well.}
  \end{subfigure}
  \caption{S-wave component of the potential as a function of the scattering energy $k$. $V_0(k)$ is in units of $\hbar^2 R^{D-2}/m$, $\gamma$ is in units of $\hbar^2/(mR^2)$, and $k$ is in units of $1/R$. The regularizer for the $D=2$ case is chosen as $\Lambda = 1/R$. The solid line is the exact s-wave component, Eq.~(\ref{eq:swave-def}), and the dashed line is the approximation obtained with the on-shell Eq.~(\ref{eq:osa}).}
  \label{fig:swk}
\end{figure}
\begin{figure}[h]
  \centering
   \includegraphics[width=\textwidth]{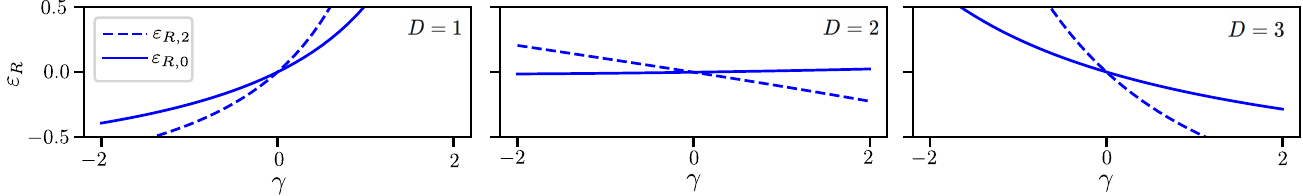}
  \caption{Relative error $\varepsilon_{R, 0}$ and $\varepsilon_{R, 2}$ on the approximation of the low-momentum expansion coefficients of the Fourier transform using the on-shell approximation. $\varepsilon_{R, 0}$ is the relative error on the value of $\tilde{g}_0=g_0$, and $\varepsilon_{R, 2}$ is the relative error on $\tilde{g}_2=g_2/2$. $\gamma$ is in units of $\hbar^2/(mR^2)$. The regularizer for the $D=2$ case is chosen as $\Lambda = 1/R$.}
  \label{fig:swe}
\end{figure}

\hspace{0.2cm}%
The general $D$-dimensional square well potential is given in coordinate space as
\begin{equation}
V(r) = \begin{cases}
\gamma, & r \leq R \,, \\
0, & r > R \,, 
\end{cases}
\end{equation}
where $\mathbf{r}\in \mathbb{R}^D$, and $r=|\mathbf{r}|$. It simply represents an attractive ($\gamma<0$) or repulsive ($\gamma>0$) potential acting when the atoms are closer than $R$.
The s-wave components for this potential are the following
\begin{equation}\label{eq:sw-swave}
    V_0(k) = \begin{cases} \displaystyle
    \gamma \ \frac{k R+\cos(k R)\sin(k R)}{k} \,, \qquad &D=1\\[1.5ex]\displaystyle
    \pi R^2 \ \gamma \ (J_0(k R)^2+J_1(k R)^2) \,, \qquad &D=2\\[1.5ex]\displaystyle
    2\pi \gamma \ \frac{k R-\cos(k R)\sin(k R )}{k^3} \,, \qquad &D=3
    \end{cases}
\end{equation}

\hspace{0.2cm}%
We solve the Schr\"odinger equation with the energy $E = \hbar^2 k^2/m$ linked to the relative momentum $k$. We also define the auxiliary momentum $k_1 = \sqrt{k^2 - \frac{m\gamma}{\hbar^2}}$.
The determination of the phase shift is done by requiring the continuity of the wavefunction $\psi(\mathbf{r})$ and its derivative at $r=R$. This implies the continuity of the logarithmic derivative $\frac{d}{dr}\log\psi(\mathbf{r}) = \psi'(\mathbf{r})/\psi(\mathbf{r})$, which is sufficient to determine the phase shift.

\hspace{0.2cm}%
Starting from the $D=1$ case, we notice that, unlike $D=2, \ 3$ we are in absence of an angular variable, and the Schr\"odinger equation is better considered in the variable $x$ over all the real axis. In this way $r=|x|$.
The parity-symmetric solution is expressed as
\begin{align}\label{eq:1dpsi}
\psi(x) &=
\begin{cases}
A \cos(k_1 x), & |x| < R \\
B\cos(kx + \operatorname{sgn}(x) \delta_0), & |x| > R
\end{cases}
\end{align}
where $\operatorname{sgn}(x)$ is the sign function of $x$.
The continuity condition at $x = R$ gives the s-wave phase shift as
\begin{equation}\label{eq:deltaws1}
\delta_0(k) = \arctan\left[\frac{k_1}{k} \tan\left( k_1 R \right)\right] - k R \,.
\end{equation}
In $D=2$, the solution to the s-wave radial Schr\"odinger equation is
\begin{align}\label{eq:2dpsi}
\psi(r) &=
\begin{cases}
A J_0(k_1 r), & r \leq R \\
B\left[ \cos\delta_0 \, J_0(k r) - \sin\delta_0 \, Y_0(k r) \right], & r > R
\end{cases}
\end{align}
and the matching condition at $r=R$ yields
\begin{equation}\label{eq:deltaws2}
\delta_0(k) =\arctan\left[
\frac{k_1 \ J_0(kR) J_0'(k_1 R) - k \ J_0(k_1 R) J_0'(kR)}
{k \ J_0(k_1 R) Y_0'(kR) - k_1 \ Y_0(kR) J_0'(k_1 R)}\right].
\end{equation}

\hspace{0.2cm}%
In $D=3$ the radial Schrödinger equation for the $l=0$ partial wave is solved under the change of variable $u(r)=r\psi(r)$, with solutions
\begin{align}\label{eq:3dpsi}
u(r) &=
\begin{cases}
A \sin(k_1 r), & r \leq R \\
B\sin(k r + \delta_0), & r > R
\end{cases}
\end{align}
again, the matching at $r=R$ leads to the phase shift
\begin{equation}\label{eq:deltaws3}
\delta_0(k) = \arctan\left[\frac{k}{k_1} \tan\left( k_1 R \right)\right] - k R \,.
\end{equation}

\hspace{0.2cm}%
The comparison of the true value of the s-wave interaction potential $V_0(k)$ and the OS approximation, as a function of $k$, is performed in Fig.~\ref{fig:swk}. The approximation is computed utilizing the exact phase shifts in Eq.~(\ref{eq:deltaws1}, \ref{eq:deltaws2}, \ref{eq:deltaws3}), combined with Eq.~(\ref{eq:osa}). In the latter relation, for the case $D=2$, the cutoff parameter is fixed to a convenient value of $\Lambda=1/R$. The approximation is seen to hold better for high values of the momentum, and for low values of $\gamma$, a similar regime of validity as the Born approximation (see also Appendix B).
Focusing into the low-energy regime, we notice that the approximation systematically overestimates (in $D=3$) or underestimates ($D=1, 2$) the low-momentum values of $V_0(k)$, irrespectively of the sign of the potential. 
Also, we evaluate the relative error in the approximation of the low-momentum expansion coefficients  $\tilde{g}_0$ and $\tilde{g}_2$ of the Fourier-transformed potential $\tilde{V}$, as represented in Fig.~\ref{fig:swe}. The relative error is low only for weak potentials, and the dependence on the dimensionality is quite marked.

\hspace{0.2cm}%
Quite remarkably, we find that the low $\gamma$ behavior of the OS approximation exactly follows the value of $V_0(k)$ for every value of momentum.
In fact, it is easy to verify that in the limit of $\gamma\to0$, the leading order of the on-shell approximation given by the substitution of Eq.~(\ref{eq:deltaws1}, \ref{eq:deltaws2}, \ref{eq:deltaws3}) into Eq.~(\ref{eq:osa})  coincide with the exact result, which is a linear function of $\gamma$ due to the linearity of Fourier and Hankel transforms. This is valid irrespectively of the fact that the potential is attractive or repulsive, and holds in every dimension.

\subsection{Delta shell}
\begin{figure}[htb]
  \centering
  \begin{subfigure}[b]{0.42\textwidth}
    \includegraphics[width=\textwidth]{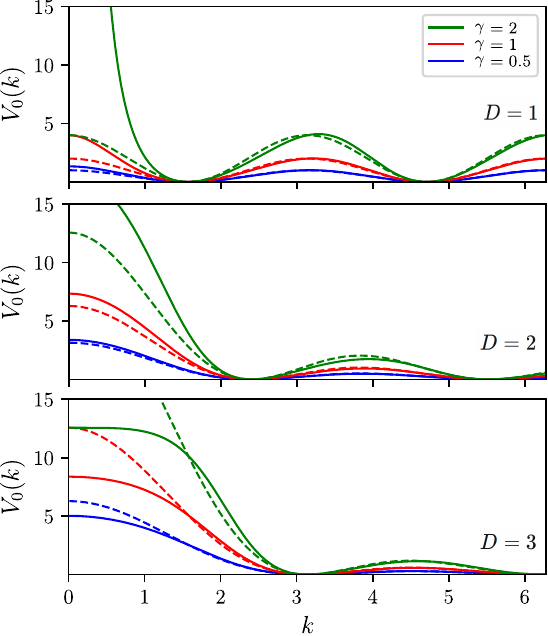}
    \caption{Attractive delta shell.}
  \end{subfigure}
  \hfill
  \begin{subfigure}[b]{0.42\textwidth}
    \includegraphics[width=\textwidth]{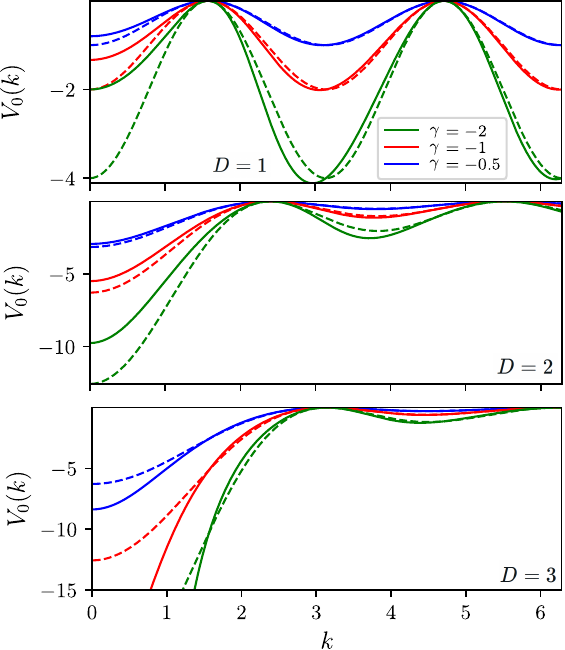}
    \caption{Repulsive delta shell.}
  \end{subfigure}
  \caption{
  On-shell s-wave component of the potential as a function of the scattering energy $k$. $V_0(k)$ is in units of $\hbar^2R^{D-2}/m$, $\gamma$ is in units of $\hbar^2/(mR)$, and $k$ is in units of $1/R$. The regularizer for the $D=2$ case is chosen as $\Lambda = 1/R$. The solid line is the exact s-wave component $V_0(k)$, and the dashed line is the approximation obtained with the on-shell approximation (\ref{eq:osa}). 
  }
  \label{fig:dsk}
\end{figure}

\begin{figure}[htb]
  \centering
   \includegraphics[width=1\textwidth]{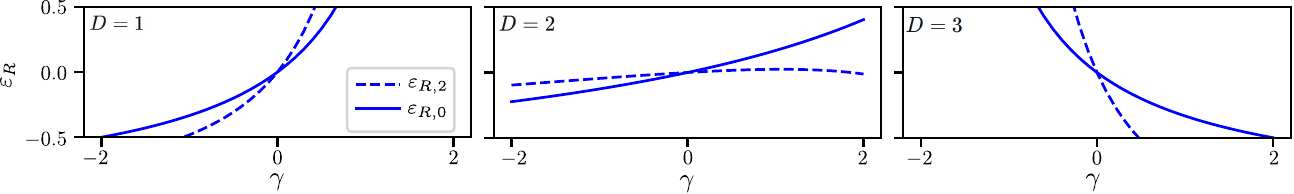}
  \caption{Relative error $\varepsilon_{R, 0}$ and $\varepsilon_{R, 2}$ on the approximation of the low-momentum expansion coefficients of the Fourier transform using the on-shell approximation. $\varepsilon_{R, 0}$ is the relative error on the value of $\tilde{g}_0=g_0$, and $\varepsilon_{R, 2}$ is the relative error on $\tilde{g}_2=g_2/2$. Parameters and units are the same as in Fig.~\ref{fig:swe}.}
  \label{fig:dse}
\end{figure}

\hspace{0.2cm}%
The general $D$-dimensional delta shell potential is given by
\begin{equation}
V(r) = \gamma \, \delta^{(1)}(R-r) \,,
\end{equation}
where $\mathbf{r}\in \mathbb{R}^D$, and $r=|\mathbf{r}|$. It represents a hard shell at a distance $R$ from the center of mass. Notice that the dimensions of $\gamma$ are independent on the dimension and are of energy times length ($\textsf{E}\textsf{L}$).
The s-wave components for this potential are the following
\begin{equation}\label{eq:ds-swave}
    V_0(k) = \begin{cases} \displaystyle
    2 \gamma \ (\cos(k R))^2\,, \qquad &D=1\\[1.5ex]\displaystyle
    2\pi R  \ \gamma \ (J_0(k R))^2 \,, \qquad &D=2\\[1.5ex]\displaystyle
     4\pi^2 R^2 \gamma \ (j_0(k R))^2 \,, \qquad &D=3
    \end{cases}
\end{equation}

\hspace{0.2cm}%
In solving the Schr\"odinger equation with the energy $E = \hbar^2 k^2/m$, we make use of the \Ra{condition fixing the discontinuity} for the wavefunction derivative at $r=R$ \Ra{as a function of $\gamma$}. We also impose the continuity of the wavefunction at $r=R$. Once again, the conditions can be combined in the form of $\frac{d}{dr}\log\psi(\mathbf{r}) = \psi'(\mathbf{r})/\psi(\mathbf{r})$, namely, as $\psi'(R^+)-\psi'(R^-)=\frac{m}{\hbar^2}\gamma \ \psi(R)$.
The solutions are analogous to the ones reported for the square well, viz. Eqs.~(\ref{eq:1dpsi}, \ref{eq:2dpsi}, \ref{eq:3dpsi}) except for the fact that $k_1$ is substituted with $k$, being that the potential in the region $r<R$ is zero in this case.
By applying the jump condition and continuity, we obtain in $D=1$
\begin{equation}\label{eq:deltads1}
\delta_0(k) = \arctan\left[-\frac{m}{\hbar^2}\frac{\gamma}{k}+\tan(k R)\right]-k R \,.
\end{equation}
In $D=2$ we obtain 
\begin{equation}\label{eq:deltads2}
    \delta_0(k) = \arctan\left[\frac{\frac{m}{\hbar^2}\gamma \ J_0(k R)^2}{\frac{m}{\hbar^2}\gamma J_0(k R)Y_0(k R)-\frac{2}{\pi R}}\right] \,,
\end{equation}
where we used the Wronksian relation $\mathcal{W}[J_0, Y_0](z)=J_0'(z)Y_0(z)-J_0(z)Y_0'(z)=2/(\pi z)$.

Finally, in $D=3$, the result is, as expected, similar to the $D=1$:
\begin{equation}\label{eq:deltads3}
    \delta_0(k)=\arctan\left[\frac{k}{k\cot(k R)+\frac{m}{\hbar^2}\gamma}\right]-k R
\end{equation}

\hspace{0.2cm}%
The comparison between $V_0(k)$ and its OS approximation as a function of $k$ is shown in Fig.~\ref{fig:dsk}. The value of the OS approximation is computed utilizing the exact phase shifts in Eq.~(\ref{eq:deltads1}, \ref{eq:deltads2}, \ref{eq:deltads3}), combined with Eq.~(\ref{eq:osa}). In $D=2$ we choose a cutoff parameter of $\Lambda=1/R$. The low-momentum regime is systematically overestimated (in $D=3$) or underestimated ($D=1, 2$), irrespectively of the sign of the potential, as in the case of the square well.
The error in the approximation of the low-momentum expansion coefficients  $\tilde{g}_0$ and $\tilde{g}_2$ of the Fourier-transformed potential $\tilde{V}$ are represented in Fig.~\ref{fig:dse}, where it is shown that such error is low only for weak potentials, and dependends strongly on the dimensionality involved.

\hspace{0.2cm}%
Remarkably, as in the case of the square well, even in the case of the delta shell potential it is easy to verify that the leading order in the low $\gamma$ behavior of the OS approximation exactly matches the value of $V_0(k)$ for every value of momentum.

\hspace{0.2cm}%
\Ra{
We remark that the dependence on the cutoff momentum $\Lambda$ of the OS approximation is essential due to the need to regularize divergences appearing in the approximation procedure. However, as reported in in Ref.~\cite{lorenzi_-shell_2023}, this dependence on $\Lambda$ has been shown  to be inessential for the determination of the equation of state of a two-dimensional Bose gas with an EFT approach, where the limit of $\Lambda \to \infty$ was used to retrieve the  pressure arising from a effective-range corrected field theory. 
The $D=2$ results illustrated in the present work show a qualitative accuracy of the OS approximation, taken using a reasonable value of the cutoff $\Lambda$ compared to the physical scales of the potential.}
\section{Conclusions}
\hspace{0.2cm}%
In this work, we assessed the validity of the on-shell (OS) approximation by directly comparing the s-wave component of the interaction potential with its estimate derived from the s-wave phase shift, focusing on square well and delta shell potentials. The approximation is found to be accurate in the weak-potential regime, with increasing precision at higher momenta. Remarkably, in the limit of weak interactions, the OS approximation reproduces the exact s-wave potential to leading order across all momenta.
These results indicate that the OS approximation yields the correct asymptotic behavior of the s-wave potential for low interaction strength, making it a reliable reference for evaluating more sophisticated approximations in this regime.
In the context of ultracold bosons, our approach is complementary to existing methods that relate phase shifts to energy shifts in systems where an external potential is also present \cite{fukuda_energy_1956, yu_calculating_2011, collin_energy-dependent_2007}. Instead, it aligns more closely with methods used in effective field theory \cite{braaten_scattering_2008}, where the goal is to construct effective interaction potentials that reproduce low-energy scattering properties, rather than compute finite-size energy corrections.
Extending the OS approximation to finite-temperature many-body systems involves additional subtleties and is a promising direction for future work.

\medskip
\textbf{Acknowledgements}
\hspace{0.2cm}%
LS acknowledges the BIRD Project “Ultra-cold atoms in curved geometries” of the University of Padova. LS is partially supported by the European Union NextGenerationEU within the National Center for HPC, Big Data, and Quantum Computing [Project No. CN00000013, CN1 Spoke 10: Quantum Computing] and by the European Quantum Flagship Project PASQuanS 2. LS acknowledges Iniziativa Specifica Quantum of Istituto Nazionale di Fisica Nucleare, the Project “Frontiere Quantistiche” within the 2023 funding programme ‘Dipartimenti di Eccellenza’ of the Italian Ministry for Universities and Research, and the PRIN 11 2022 Project “Quantum Atomic Mixtures: Droplets, Topological Structures, and Vortices”.

\subsubsection*{Data availability}
\hspace{0.2cm}%
No new data was generated in the present work.

\subsubsection*{Conflict of interest}
\hspace{0.2cm}%
The author declare no conflict of interest.

\section*{Appendix A: Fourier transform in presence of central symmetries}
\hspace{0.2cm}%
The symmetry of the interaction potential can be used to simplify the calculations of the Fourier transform.
While in the $D=1$ this is trivial, namely we consider a Fourier integral of an even-parity function, in $D=2$ and $D=3$ the circular and spherical symmetry can be used to reduce the double and triple Fourier integrals to a single integral, constituting the Hankel and Spherical Hankel transforms. The standard $D$-dimensional Fourier transform is defined as
\begin{equation}
\tilde{f}(\mathbf{q}) = \int d^D\mathbf{r} \,  f(\mathbf{r}) e^{-i \mathbf{q} \cdot \mathbf{r}} \,.
\label{eq:fourier-Dd}
\end{equation}

In 1D we have the relation
\begin{equation}\label{eq:spherift-1d}
    \tilde{f}(q) = 2 \int_0^{\infty} dx \ f(r) \cos(qx) \,.
\end{equation}

In 2D, with circular symmetry, i.e.\ $ f(\mathbf{r}) = f(r) $, we have also $ \tilde{f}(\mathbf{q}) = \tilde{f}(q) $. 
In polar coordinates, the Fourier integral Eq.~(\ref{eq:fourier-Dd}) with $D=2$ reads
\begin{align}
    \tilde{f}(q) &= \int d^2\mathbf{r}\, f(r)\, e^{-i \mathbf{q} \cdot \mathbf{r}} 
    = \int_0^{2\pi} d\theta \int_0^\infty dr\, r\, f(r)\, e^{-iqr\cos\theta} \,.
\end{align}
Evaluating the angular integral, we use the identity
\begin{align}
    \int_0^{2\pi} d\theta\, e^{-iqr\cos\theta} = 2\pi J_0(qr), 
\end{align}
that defines the zeroth-order Bessel function $J_0$ \cite{abramowitz_handbook_1965, watson_treatise_1995}. The final form of the transform is therefore
\begin{align}\label{eq:spherift-2d}
    \tilde{f}(q) = 2\pi \int_0^\infty dr\, r\, f(r) J_0(qr),
\end{align}
and it is also known as Hankel transform of order zero \cite{siegman_quasi_1977}, that we denote as $\mathcal{H}^{2D}_0[f(r)](q)$.
The inverse transform follows a similar argument, for which it is shown that the transform is involutive up to a multiplicative constant.
In 3D, we assume spherical symmetry, and similarly to the 2D case $ f(\mathbf{r}) = f(r) $, implying $ \tilde{f}(\mathbf{q}) = \tilde{f}(q) $. Using to spherical coordinates to express the Eq.~(\ref{eq:fourier-Dd}), %
we get:
\begin{equation}
\tilde{f}(q) = 2\pi \int_0^\infty dr \, r^2 f(r) \int_0^\pi d\theta \, \sin\theta \, e^{-i q r \cos\theta} \,,
\label{eq:fourier-angular}
\end{equation}
Evaluating the angular integrals we recover the structure of the spherical Bessel function (also known as Riccati-Bessel function) \cite{rakityansky_jost_2023, newton_scattering_2013}
\begin{equation}
\int_0^\pi d\theta \, \sin\theta \, e^{-i q r \cos\theta} = \frac{2 \sin(q r)}{q r} = 2 j_0(qr)  \,, \label{eq:theta-integral}
\end{equation}
So we rewrite the Fourier transform as:
\begin{equation}\label{eq:spherift-3d}
\tilde{f}(q) = 4\pi \int_0^\infty dr \, r^2 f(r) j_0(q r) \,.
\end{equation}
This is also known as the spherical Hankel transform of order zero \cite{abramowitz_handbook_1965}, denoted by $\mathcal{H}^{3D}_0[f(r)](q)$.
Discussion of the inverse transforms is completely analogous to the 2D case.
We remark the relation between the spherical Bessel and Bessel functions: 
\begin{equation}
    j_n(x) = \sqrt{\frac{\pi}{2x}} J_{n + \frac{1}{2}}(x) \,,
\end{equation}
from which it is seen that the spherical Hankel and Hankel transforms are related as $\mathcal{H}^{3D}_0[f(r)](q) = (2\pi)^{3/2}/q^{1/2} \; \mathcal{H}^{2D}_{1/2}[\sqrt{r}f(r)](q)$.

\hspace{0.2cm}%
A useful set of relations is the expansion of a plane wave in partial waves. In $D=1$ it is particularly simple, taking the form of Euler formula $e^{ikx}=\cos(kx)+i\sin(kx)$. In $D=2$, it reads:
\begin{equation}
e^{i \mathbf{k} \cdot \mathbf{r}} = \sum_{m=-\infty}^\infty i^m J_m(k r) e^{i m \phi} \,.
\end{equation}
In $D=3$ it is 
\begin{equation}
e^{i \mathbf{k} \cdot \mathbf{r}} = \sum_{\ell=0}^\infty i^\ell (2\ell + 1) j_\ell(k r)  P_\ell(\cos \phi) \,,
\end{equation}
with $j_\ell$ the spherical Bessel function of order $\ell$, and $P_\ell$ the $\ell$-th Legendre polynomial. 

\section*{Appendix B: Born approximation}
\hspace{0.2cm}%
We recall the Born approximation for the s-wave scattering. The relation reads $T_0(k)\approx V_0(k)$, where $T_0(k)$ is the on-shell s-wave $T$ matrix, and $V_0(k)$ the s-wave potential. To derive the Born approximation, the off-shell components of both $T$-matrix and interaction potential is completely neglected, as in the OS method. The Born approximation is the first-order solution of the so-called Born series \cite{sakurai_modern_1985}. By using the explicit expressions that define the $T$-matrix in terms of the s-wave phase shift, as found in Ref.~\cite{lorenzi_-shell_2023}, the relation $T_0(k)\approx V_0(k)$ reads
\begin{equation}\label{eq:born}
     V_0(k) \overset{\mathrm{B}}{=} \begin{cases}
     -\frac{2\hbar^2}{m} k\  \mathcal{C}(k) \,, \qquad &D=1 \\
     -\frac{4\hbar^2}{m} \mathcal{C}(k)\,, \qquad  &D=2 \\ 
     -\frac{4\pi\hbar^2}{m} k^{-1} \ \mathcal{C}(k)  \,, \qquad &D=3\\
     \end{cases}
\end{equation}
where $\mathcal{C}(k) = (\cot(\delta_0(k))-i)^{-1}$, and $\mathrm{B}$ indicates the use of Born approximation. We remark that this approximation is expected to be valid for weak potentials, and it is utilized for $k\to0$ in the treatment of contact interaction. However, its general form reported here, manifest a nonzero imaginary part, that is unphysical for real and centrosymmetric potentials. 


\begin{thebibliography}{10}
\providecommand{\url}[1]{\texttt{#1}}
\providecommand{\urlprefix}{URL }

\bibitem{stoof_ultracold_2010}
H.~T.~C. Stoof, D.~B.~M. Dickerscheid, K.~Gubbels,
\newblock \emph{Ultracold {Quantum} {Fields}},
\newblock Springer Netherlands, \textbf{2010}.

\bibitem{pethick_boseeinstein_2008}
C.~J. Pethick, H.~Smith,
\newblock \emph{Bose–{Einstein} {Condensation} in {Dilute} {Gases}},
\newblock Cambridge University Press, \textbf{2008}.

\bibitem{khaykovich2002formation}
L.~Khaykovich, F.~Schreck, G.~Ferrari, T.~Bourdel, J.~Cubizolles, L.~D. Carr,
  Y.~Castin, C.~Salomon,
\newblock \emph{Science} \textbf{2002}, \emph{296}, 5571 1290.

\bibitem{bardin_quantum_2024}
A.~Bardin, F.~Lorenzi, L.~Salasnich,
\newblock \emph{New J. Phys.} \textbf{2024}, \emph{26}, 1 013021.

\bibitem{adhikari_quantum_1986}
S.~K. Adhikari,
\newblock \emph{Am. J. Phys.} \textbf{1986}, \emph{54}, 4 362.

\bibitem{barlette2000quantum}
V.~E. Barlette, M.~M. Leite, S.~K. Adhikari,
\newblock \emph{Eur. J. Phys.} \textbf{2000}, \emph{21}, 5 435.

\bibitem{tononi_low-dimensional_2023}
A.~Tononi, L.~Salasnich,
\newblock \emph{Nat. Rev. Phys.} \textbf{2023}, \emph{5}, 7 398.

\bibitem{egorov_measurement_2013}
M.~Egorov, B.~Opanchuk, P.~Drummond, B.~V. Hall, P.~Hannaford, A.~I. Sidorov,
\newblock \emph{Phys. Rev. A} \textbf{2013}, \emph{87}, 5 053614.

\bibitem{bartenstein_precise_2005}
M.~Bartenstein, A.~Altmeyer, S.~Riedl, R.~Geursen, S.~Jochim, C.~Chin, J.~H.
  Denschlag, R.~Grimm, A.~Simoni, E.~Tiesinga, C.~J. Williams, P.~S. Julienne,
\newblock \emph{Phys. Rev. Lett.} \textbf{2005}, \emph{94}, 10 103201.

\bibitem{inouye_observation_1998}
S.~Inouye, M.~R. Andrews, J.~Stenger, H.-J. Miesner, D.~M. Stamper-Kurn,
  W.~Ketterle,
\newblock \emph{Nature} \textbf{1998}, \emph{392}, 6672 151.

\bibitem{cappellaro_thermal_2017}
A.~Cappellaro, L.~Salasnich,
\newblock \emph{Phys. Rev. A} \textbf{2017}, \emph{95}, 3 033627.

\bibitem{planasdemunt2024equation}
M.~Planasdemunt-Hospital, J.~Pera, J.~Boronat,
\newblock \emph{Phys. Rev. Research} \textbf{2024}, \emph{6}, 4 L042071.

\bibitem{roth_effective_2001}
R.~Roth, H.~Feldmeier,
\newblock \emph{Phys. Rev. A} \textbf{2001}, \emph{64}, 4 043603.

\bibitem{castin_bose-einstein_2001}
Y.~Castin,
\newblock In R.~Kaiser, C.~Westbrook, F.~David, editors, \emph{Coherent atomic
  matter waves}. Springer, Berlin, Heidelberg, \textbf{2001} 1--136.

\bibitem{lorenzi_-shell_2023}
F.~Lorenzi, A.~Bardin, L.~Salasnich,
\newblock \emph{Phys. Rev. A} \textbf{2023}, \emph{107}, 3 033325.

\bibitem{newton_scattering_2013}
R.~G. Newton,
\newblock \emph{Scattering {Theory} of {Waves} and {Particles}},
\newblock Springer Science \& Business Media, \textbf{2013}.

\bibitem{rakityansky_jost_2023}
S.~A. Rakityansky,
\newblock \emph{Jost {Functions} in {Quantum} {Mechanics}: {A} {Unified}
  {Approach} to {Scattering}, {Bound}, and {Resonant} {State} {Problems}},
\newblock Springer International Publishing, \textbf{2023}.

\bibitem{taylor_scattering_2012}
J.~R. Taylor,
\newblock \emph{Scattering {Theory}: {The} {Quantum} {Theory} of
  {Nonrelativistic} {Collisions}},
\newblock Courier Corporation, \textbf{2012}.

\bibitem{macedo-lima_scattering_2023}
M.~Macêdo-Lima, L.~Madeira,
\newblock \emph{Rev. Bras. Ensino Fís.} \textbf{2023}, \emph{45} e20230079.

\bibitem{arnecke_effective-range_2006}
F.~Arnecke, H.~Friedrich, J.~Madroñero,
\newblock \emph{Phys. Rev. A} \textbf{2006}, \emph{74}, 6 062702.

\bibitem{jeszenszki_s-wave_2018}
P.~Jeszenszki, A.~Y. Cherny, J.~Brand,
\newblock \emph{Phys. Rev. A} \textbf{2018}, \emph{97}, 4 042708.

\bibitem{fu_beyond_2003}
H.~Fu, Y.~Wang, B.~Gao,
\newblock \emph{Phys. Rev. A} \textbf{2003}, \emph{67}, 5 053612.

\bibitem{veksler_simple_2014}
H.~Veksler, S.~Fishman, W.~Ketterle,
\newblock \emph{Phys. Rev. A} \textbf{2014}, \emph{90}, 2 023620.

\bibitem{zinner_effective_2012}
N.~T. Zinner,
\newblock \emph{Journal of Atomic and Molecular Physics} \textbf{2012},
  \emph{2012}, 1 241051.

\bibitem{pendse_probing_2018}
A.~Pendse, A.~Bhattacharyay,
\newblock \emph{J. Phys.: Condens. Matter} \textbf{2018}, \emph{30}, 45 455602.

\bibitem{bethe_theory_1949}
H.~A. Bethe,
\newblock \emph{Phys. Rev.} \textbf{1949}, \emph{76}, 1 38.

\bibitem{schw}
J.~Schwinger,
\newblock \emph{Phys. Rev.} \textbf{1947}, , 72 742.

\bibitem{Adhikari_2018}
S.~K. Adhikari,
\newblock \emph{Eur. J. Phys.} \textbf{2018}, \emph{39}, 5 055403.

\bibitem{sakurai_modern_1985}
J.~J. Sakurai, S.~F. Tuan,
\newblock \emph{Modern {Quantum} {Mechanics}},
\newblock Benjamin/Cummings Pub., \textbf{1985}.

\bibitem{adhikari_effective_1982}
S.~K. Adhikari, J.~A. Torreão,
\newblock \emph{Phys. Lett. B} \textbf{1982}, \emph{119}, 4 245.

\bibitem{barlette2001integral}
V.~E. Barlette, M.~M. Leite, S.~K. Adhikari,
\newblock \emph{Am. J. Phys.} \textbf{2001}, \emph{69}, 9 1010.

\bibitem{abramowitz_handbook_1965}
M.~Abramowitz, I.~A. Stegun,
\newblock \emph{Handbook of {Mathematical} {Functions}: {With} {Formulas},
  {Graphs}, and {Mathematical} {Tables}},
\newblock Courier Corporation, \textbf{1965}.

\bibitem{gradshteyn_table_2014}
I.~S. Gradshteyn, I.~M. Ryzhik,
\newblock \emph{Table of {Integrals}, {Series}, and {Products}},
\newblock Academic Press, \textbf{2014}.

\bibitem{fukuda_energy_1956}
N.~Fukuda, R.~G. Newton,
\newblock \emph{Phys. Rev.} \textbf{1956}, \emph{103}, 5 1558.

\bibitem{yu_calculating_2011}
Z.~Yu, G.~Baym, C.~J. Pethick,
\newblock \emph{J. Phys. B: At. Mol. Opt. Phys.} \textbf{2011}, \emph{44}, 19
  195207.

\bibitem{collin_energy-dependent_2007}
A.~Collin, P.~Massignan, C.~J. Pethick,
\newblock \emph{Phys. Rev. A} \textbf{2007}, \emph{75}, 1 013615.

\bibitem{braaten_scattering_2008}
E.~Braaten, M.~Kusunoki, D.~Zhang,
\newblock \emph{Ann. Phys.} \textbf{2008}, \emph{323}, 7 1770.

\bibitem{watson_treatise_1995}
G.~N. Watson,
\newblock \emph{A {Treatise} on the {Theory} of {Bessel} {Functions}},
\newblock Cambridge University Press, \textbf{1995}.

\bibitem{siegman_quasi_1977}
A.~E. Siegman,
\newblock \emph{Opt. Lett., OL} \textbf{1977}, \emph{1}, 1 13.

\end{thebibliography}

\end{document}